\documentclass[fleqn,10pt]{wlscirep}
\usepackage[utf8]{inputenc}
\usepackage[T1]{fontenc}
\usepackage{upgreek}
\usepackage[version=4]{mhchem}
\usepackage{mathtools}

\title{3D Imaging of directional multi-scale cellulose nanostructures through multi-directional dark-field neutron tomography}

\author[1*+]{Matteo Busi}
\author[1,2*+]{Elisabetta Nocerino}
\author[2]{Agnes $\rm{\mathbf{\mathring{A}}}$hl}
\author[2]{Lennart Bergström}
\author[1,*]{Markus Strobl}
\affil[1]{Applied Materials Group, Paul Scherrer Institut, 5232 Villigen, Switzerland}
\affil[2]{Department of Materials and Environmental Chemistry, Stockholm University, SE-10691 Stockholm, Sweden}

\affil[*]{matteo.busi@psi.ch; elisabetta.nocerino@mmk.su.se; markus.strobl@psi.ch}
\affil[+]{these authors contributed equally to this work}

\begin{abstract}
Hierarchical biomaterials embody nature's intricate design principles, offering advanced functionalities through the complex, multi-level organization of their molecular and nanosized building blocks. However, the comprehensive characterization of their 3D structure remains a challenge, particularly due to radiation damage caused by conventional X-ray- and electron-based imaging techniques, as well as due to the length scale limitations of scattering-based investigation methods. Here, we present a study utilizing multi-directional dark-field neutron imaging in tomographic mode to visualize the 3D nanoarchitecture of nanocellulose solid foams, a class of sustainable materials possessing complex and highly tunable hierarchical structures. By exploiting the unique properties of neutrons as a probe, this non-destructive method circumvents the inherent limitations of damage-inducing ionizing radiation, preserving the structural and chemical integrity of the biomaterials, and allowing for truly multiscale characterization of the spatial orientation and distribution of cellulose nano fibrils within large-volume samples. In particular, the study showcases the 3-dimensional anisotropic orientation and degree of alignment of nanofibrils with different crystallinity, across various length scales, from nanometers to centimeters. This approach offers a valuable and generally applicable tool for multi-scale characterisation of biobased materials where complex nanoscale arrangements inform macroscopic properties.

\end{abstract}

\begin{document}
\flushbottom
\maketitle

\thispagestyle{empty}
\section{Main}
Hierarchical structures in nature offer a profound blueprint for resilience, efficiency, and beauty, seamlessly weaving the fabric of life from the microscopic to the astronomical scale. Each level of the hierarchy supports and is supported by the others, creating robust and flexible structures that can be found in many sophisticated natural systems~\cite{national1994hierarchical}. The organization of cells in living beings, the structure of bones and tendons, the branching of trees, or the complex ecosystems of forests, oceans, and deserts exemplify nature's ability in building such sustainable and adaptable systems, that have inspired scientific and technological advancements for centuries, leading to innovations that try to mimic their properties.

In the rapidly evolving landscape of materials science, the pursuit of sustainable and high-performance materials has led to significant interest in extreme aspect ratio (length-to-width ) materials, including nanocellulose fibrils~\cite{nechyporchuk2016production}, carbon nanotubes~\cite{yanagi2018intersubband}, and certain polymeric nanostructures such as spider silk~\cite{andersson2017biomimetic}. A prime instance of exploiting extreme aspect ratio materials for functionalization by self-assembly of nanostructures is found in nanocellulose-based foams~\cite{kriechbaum2018analysis,moon2011cellulose,lavoine2017nanocellulose}, which serve as exemplary study cases within the framework of complex hierarchical systems. These porous materials, derived from abundant and renewable resources, display remarkable properties that make them suitable for various applications, ranging from lightweight structural components to thermal insulation and beyond~\cite{ferreira2020porous,chen2023lightweight,wicklein2015thermally,li2023cellulose,apostolopoulou2021thermally,apostolopoulou2021effect,khalil2012green}. However, clarifying the fundamental working principles of nanocellulose foams, to unlock their full potential, is contingent on the profound understanding of their hierarchical nanostructured architecture, which relies on the specific arrangement and orientation of their elementary building blocks (i.e., the nanocellulose fibrils) across different length scales. These are general characteristics of condensed matter, in which structure and physical properties are always deeply intertwined \cite{nocerino2022comprehensive,nocerino2023multiple,nocerino2023competition,nocerino2023q,nocerino2024ion}. Materials with extreme aspect ratios exhibit significant anisotropy in their physical properties, reflecting their structural anisotropy, and understanding how these properties are influenced by the alignment of their fibrous constituents in complex arrangements is crucial, especially for applications requiring specific directional performance \cite{goh2014directional,zhang2005aligned,griffin2022scalable}.

Traditional structural characterization techniques, though invaluable, frequently prove inadequate in capturing the multiscale complexity of these materials. This is because they are typically optimized for a specific length scale (nano, micro, or macro, but not across multiple scales simultaneously~\cite{nygaard2024formax}), or require a compromise between resolvable length scales and sample size, meaning that, in order to resolve nanometer-scale features, the sample needs to have sub-millimeter to millimeter dimensions \cite{liebi2015nanostructure,sanchez2021x,azevedo2024connectomic}. These limitations hold also for new emerging technologies, such as volume electron microscopy \cite{peddie2022volume} as well as for the recently developed upgrades on phase contrast X-ray methods \cite{kagias2019diffractive,wirtensohn2024nanoscale}. It is within this context that neutron dark-field imaging (NDFI)~\cite{strobl2008neutron} comes forth as a radical advancement, offering unprecedented insights into the 3D nanostructure of nanocellulose foams, with a clear potential to extend its application to many key materials with hierarchically organized structures, natural or artificial.

Unlike traditional radiographic imaging methods that depend on attenuation through macroscopic structures, NDFI utilizes the scattering contrast from structures on the nanometer to micrometer scale, thus, providing sensitivity to microstructural variations within the material \cite{strobl2014general,strobl2015quantitative,strobl2017small}. Here, the latest development in the framework of NDFI, the multi-directional NDFI approach~\cite{busi2023multi,valsecchi2019visualization,valsecchi2020characterization}, is noteworthy as it enables to probe anisotropic local small angle scattering and, thus, anisotropic nanostructures and their preferential orientation depending on the location in a bulk sample.


By leveraging the utilization of neutrons as probe, NDFI allows for the characterization of samples in their native state, circumventing the need for sectioning or altering the specimen, thereby preserving both their structural and chemical integrity. This aspect represents a significant advantage with respect to the more widespread X-ray- and electron-based structural characterization methods which, beyond their aforementioned lengthscale and resolution limitations compelling millimeter- and micrometer-sized samples for nanometer-scale investigations, can also induce significant radiation damage on the probed material~\cite{howells2009assessment,beetz2003soft}. The latter is a general, well-known, and severely limiting issue associated to the fundamental interactions between matter and electromagnetic radiation~\cite{holton2009beginner,olsson2022multiscale}. As X-rays deposit energy in the measured material, generating electron cascades that can break chemical bonds, induce redox processes and create free radicals, the sample's structure and chemistry is altered. This problem scales with photon exposure (i.e., radiation dose) but is also related to the X-ray energy, hence, to the specific measurement conditions and to the X-ray absorption cross sections of the scattering centers. Such alterations ultimately lead to the destruction of the measured specimen through a complex, irreversible, and unavoidable process which occurs while measuring~\cite{meents2010origin}. Electron-based structural characterization methods, such as electron microscopy and scattering, face similar limitations \cite{glaeser1978radiation,glaeser1971limitations}.

Addressing the issue of radiation damage is particularly relevant when measuring sensitive and fragile porous materials such as nanocellulose foams, as well as for most bio-based specimen. Here the inherent limitations of X-ray based imaging methods become apparent, as they require high doses to allow the acquisition of high-resolution images with sufficient contrast and to compensate for significant attenuation of the X-ray beam through the samples~\cite{simons2015dark}, they mostly allow investigation of features at length scales of tens of micrometers~\cite{guo2024grating}, and may necessitate extensive and complex sample preparation involving invasive manipulation of the measured material \cite{duncan2022x}. 

NDFI overcomes these critical limitations as neutrons can easily probe the entire volume of centimeter-sized samples in real-life conditions, without inducing any relevant radiation damage even in sensitive structures such as nanocellulose. These features ensure significant and representative insights into the material’s overall architecture, from the nano to the macro scale. The aspect of observations being enabled for the full macroscopic sample volume represents also a remarkable practical advantage when measuring nanocellulose foams, which cannot be prepared in sizes smaller than a cubic centimeter, to ensure compactness and proper alignment of the nanofibrils in the produced sample material~\cite{shao2020freeze}. The foams are also highly compressible and fragile, owing to their extremely low density porous structure. Attempting to cut them to small sizes may affect their structure in uncontrollable ways, potentially compromising the integrity of the material already at the preparation stage prior the measurement.


It should be emphasized that NDFI does not directly detect neutron small angle scattering (SAS) patterns but probes directly the real space correlation functions of scattering structures. The local visibility reduction of a spatially modulated neutron beam in NDFI, which results from small-angle scattering interactions within the sample as a function of the probed autocorrelation lengths, represents a back-transformation of the scattering function from Fourier space to real space \cite{strobl2014general}. In contrast to conventional SAS, multiple scattering does not hinder the analyses and it is possible to accurately resolve the 3D orientation and arrangement of nanostructures regardless of the specific sample geometry. By way of example, in samples displaying highly symmetrical shapes (spherical- or cylinder-shaped) SAS patterns appear identical from multiple directions, making it difficult to accurately resolve their nanostructures with scanning SAS methods \cite{liebi2015nanostructure}. In addition, neutrons are sensitive to light elements such as hydrogen, which is prevalent in biological tissues and many organic compounds, and offer the unique ability to differentiate between elements that are close to each other in the periodic table (as, e.g., carbon and oxygen). X-ray and electron based techniques are significantly less effective in detecting light elements or distinguishing between elements with similar electron densities. These features, together with contrast variation through isotope exchange, make neutrons in general, and NDFI in particular, invaluable for studying biological structures, polymers, and other organic materials where these elements play a crucial role.

In this work, we demonstrate the potentials of NDFI by imaging the 3-dimensional nanostructure of as-grown cylindrical nanocellulose solid foams in their entirety ($\sim$ 5 cm height, 1 cm diameter), revealing the local alignment degree and orientation directions of fibrils within the foam walls throughout the full sample volume. This capability for truly multiscale characterization of biological samples, from the nanometer to centimeter scale, is unmatched by currently available structural characterization techniques.

\begin{figure}[ht]
    \centering
    \includegraphics[width=1\linewidth]{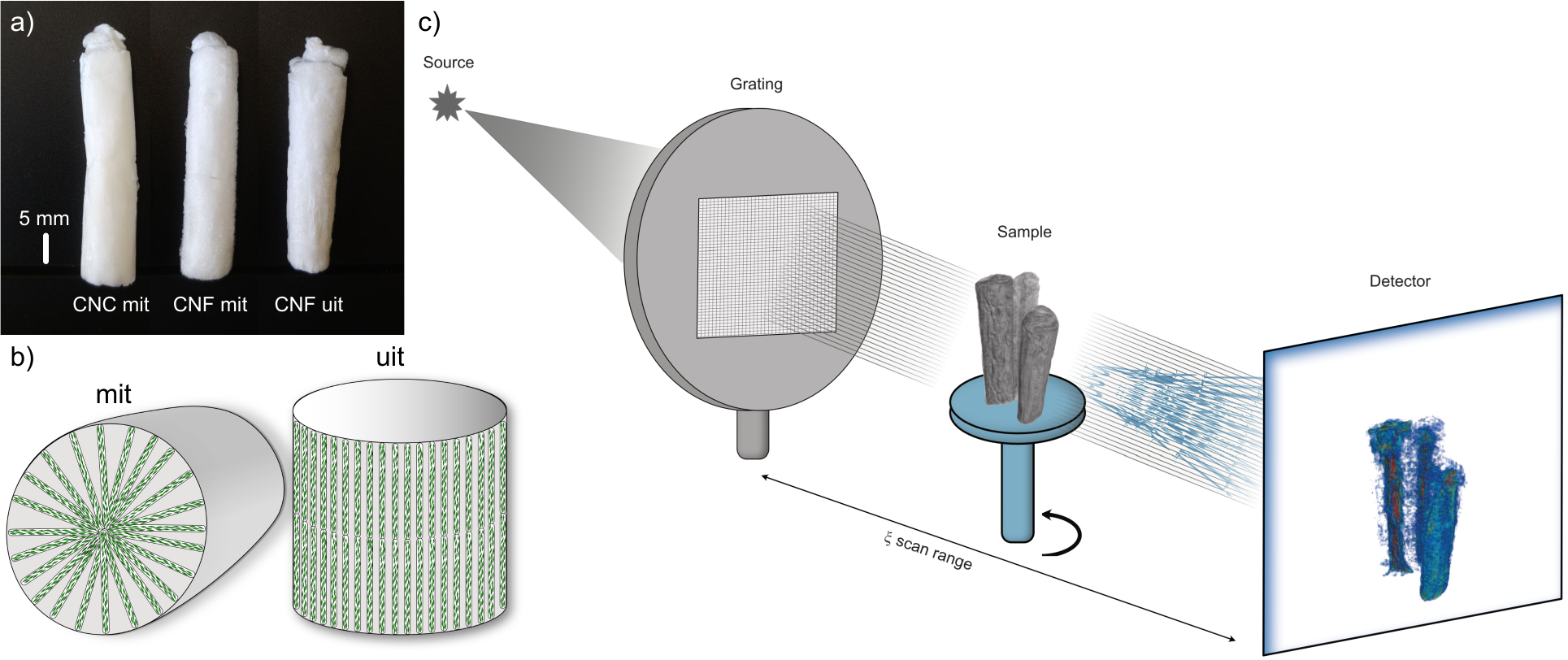}
    \caption{a) Picture of the as-grown nanocellulose foams investigated in our experiment. b) Schematic representation of the nanocellulose particles (green diamonds) aligned in the pore walls of the foams fabricated with the multi-directional ice-templating (mit) and with the uni-directional ice-templating (uit) methods. c) Schematic representation of the experimental setup for multi-directional NDFI. As the neutron beam encounters the grating, a spatially-modulated attenuation pattern according to the grating's design is created and recorded with a neutron detector. Modulations in the attenuation pattern due to the directional dependent small angle scattering interactions within the sample that are probed by NDFI provide information about anisotropy in its microstructure.}
    \label{setup}
\end{figure}

The samples presented in this work are three cylinder-shaped highly-oriented nanocellulose foams, obtained by ice-templating~\cite{fan2018creating,xu2019efficient} of nano-structured cellulose materials (CNM)~\cite{moon2011cellulose}, that can be obtained in the form of cellulose nanocrystals (CNC) or cellulose nanofibrils (CNF). Fig.~\ref{setup} a) displays the nanocellulose foams as they appear at the end of the preparation process.

Ice-templating, also known as freeze-casting, is a well-established fabrication technique used to prepare highly oriented nano-composites with controlled microstructures~\cite{shao2020freeze}. In the present work this method involves suspending nanocellulose particles in deionized water and freezing the suspension under controlled conditions. As the water freezes, ice crystals form and grow, pushing the suspended nanocellulose particles aside creating a network-like structure. During freezing, the nanocellulose particles become aligned in parallel to the temperature gradient, thus, highly ordered and tightly packed in the liquid channels between the growing ice crystals. Upon subsequent freeze-drying~\cite{fan2018creating,xu2019efficient}, the ice crystals sublimate, leaving behind a porous scaffold composed of the aligned nanocellulose particles. By carefully controlling the freezing conditions, such as temperature gradients and freezing rates, it is possible to tailor the microstructure and orientation of the resulting nanocellulose foams.

For the foams used in this work, we employed both CNC and CNF cellulose nanomaterials and applied two types of alignment (further details can be found in the Methods
section): 

\begin{itemize}
 \item the multi-directional ice-templating method (mit), in which the nanocellulose particles are radially oriented towards the center of the foam, laying in the plane perpendicular to the vertical axis of the cylinder;
 \item the uni-directional ice-templating method (uit), in which the orientation of the nanocellulose particles is parallel to the vertical axis of the cylindrical foam.
\end{itemize}

A schematic representation of the resulting principal alignment of the nanocellulose particles in the foam walls is displayed in Fig.~\ref{setup} b). The multidirectional NDFI approach in tomographic mode used in this work was proven to successfully resolve fiber orientations and alignment degree in these foams for structures on a length scale from a few tens to a few hundreds nanometers. This could be achieved by probing multiple scattering directions simultaneously in each pixel of the images of the whole centimeter-sized foam samples, thereby providing an accurate 3D multiscale structural characterization with a single measurement. Fig.~\ref{setup} c) displays a schematic view of the experimental setup.
Due to their structural anisotropy from the fibril alignment imposed during ice-templating, and their unsuitability for extensive studies under damage-inducing ionizing radiation, these systems lend themselves well to neutron dark-field imaging studies and test. 


\section*{Discussion}
\subsection{Tomographic multi-directional neutron dark-field imaging}
The working principle of the novel multi-directional NDFI technique (schematically illustrated in Fig.~\ref{setup} c) relies on the use of a single grating with an absorbing pattern that generates a spatially-modulated neutron beam with 2D periodical patterns that are directly resolved by a high-resolution neutron imaging detector~\cite{busi2023multi}. The distortions of the modulated pattern due to the interactions within the sample allow to detect and analyze anisotropies in the sample’s nanostructure.

The key advancement of this multi-directional NDFI technique over e.g. traditional Talbot-Lau interferometry, that employs three line gratings and is limited to probing a single scattering direction at a time~\cite{pfeiffer2006neutron}, lies in its ability to simultaneously probe multiple scattering directions without the need to change the orientation of the sample or the gratings. This enhances the efficiency of the imaging process, allowing for characterization of anisotropic and heterogeneous microstructures in a single measurement. Moreover, the use of a single absorption grating simplifies the experimental setup and makes the technique achromatic, thus highly compatible with and advantageous for time-of-flight measurements\cite{strobl2019achromatic}.

The raw intensity-modulated images undergo spatial harmonic analysis, where a 2D Fourier transform is applied to a defined analyzing window for each detector pixel, both for the sample and reference (without the sample) images. The Fourier image is transformed into polar coordinates to facilitate the extraction of the discrete peaks (i.e. harmonic peaks) in the spatial frequency domain. The first harmonic coefficient provides the transmission, while the second harmonic coefficients allow the calculation of the modulation visibility and directional dark-field signal. Further details about the procedure for the image processing retrieval is reported by Busi et al~\cite{busi2023multi}. The structure sizes corresponding to the probed scattering signal in the dark-field images are defined by the autocorrelation length $\xi=\lambda L_\mathrm{s}/p$, where $\lambda$ is the average neutron wavelength, $L_\mathrm{s}$ is the sample-to-detector distance and $p$ is the grating period. By varying e.g. the sample-detector distance, different autocorrelation lengths are probed, enabling the extraction of microstructural information with respect to corresponding length scales.
For this study, a hexagonal honeycomb 2D pattern grating with a period of 350 $\upmu$m was used, which enables scattering sensitivity in three directions simultaneously. The specific values of the three scattering directions depend on the orientation of the grating. For this experiment, the grating was oriented such that the scattering sensitivity of the dark-field signal would be along 0° (horizontal), 60° and 120° in the reference frame of the plane normal to the beam direction. 
The distance between the grating and the detector was fixed at 27 cm, at the peak visibility of the inverse-pattern regime~\cite{busi2022cold}.
For the tomographic acquisition, the samples were positioned at a distance of 25 cm upstream the detector, corresponding to a probed autocorrelation length of 200 nm, well suited for the investigation of the structural composition of the cellulose nanofibrils and nanocrystals.
Since for the investigated samples the principal nanostructural anisotropy of interest was between the horizontal and vertical directions, the dark-field signal along the vertical direction was approximated by interpolation of the two diagonal signals.
72 projections with sample rotation angles in steps of approximately 4.9° between 0 and 360° were acquired with an exposure of 15 minutes for each. A reference image without the sample and a dark current background image were acquired for correction and data normalization procedures. Three different sinograms were obtained based on the attenuation, on the horizontal and vertical dark-field images of the samples, and according to Beer-Lambert’s law~\cite{vontobel2006neutron}.
The tomographic reconstructions were conducted with the SIRT~\cite{palenstijn2015distributed} algorithm implemented in the ASTRA toolbox~\cite{van2016fast}.
The geometrical parameters required for reconstruction were first tuned using the attenuation images, and kept fixed for the directional dark-field tomography reconstruction. To highlight the orientation of the probed nanostructure, the eccentricity was calculated assuming an ellipsoidal model for the 2D scattering function. The eccentricity ($e$) represents the level of preferential structural alignment, ranging from horizontal ($e < 0$) via isotropic (random distribution $e = 0$) to vertical ($e > 0$).
\begin{figure}[h!]
    \centering
    \includegraphics[width=0.85\linewidth]{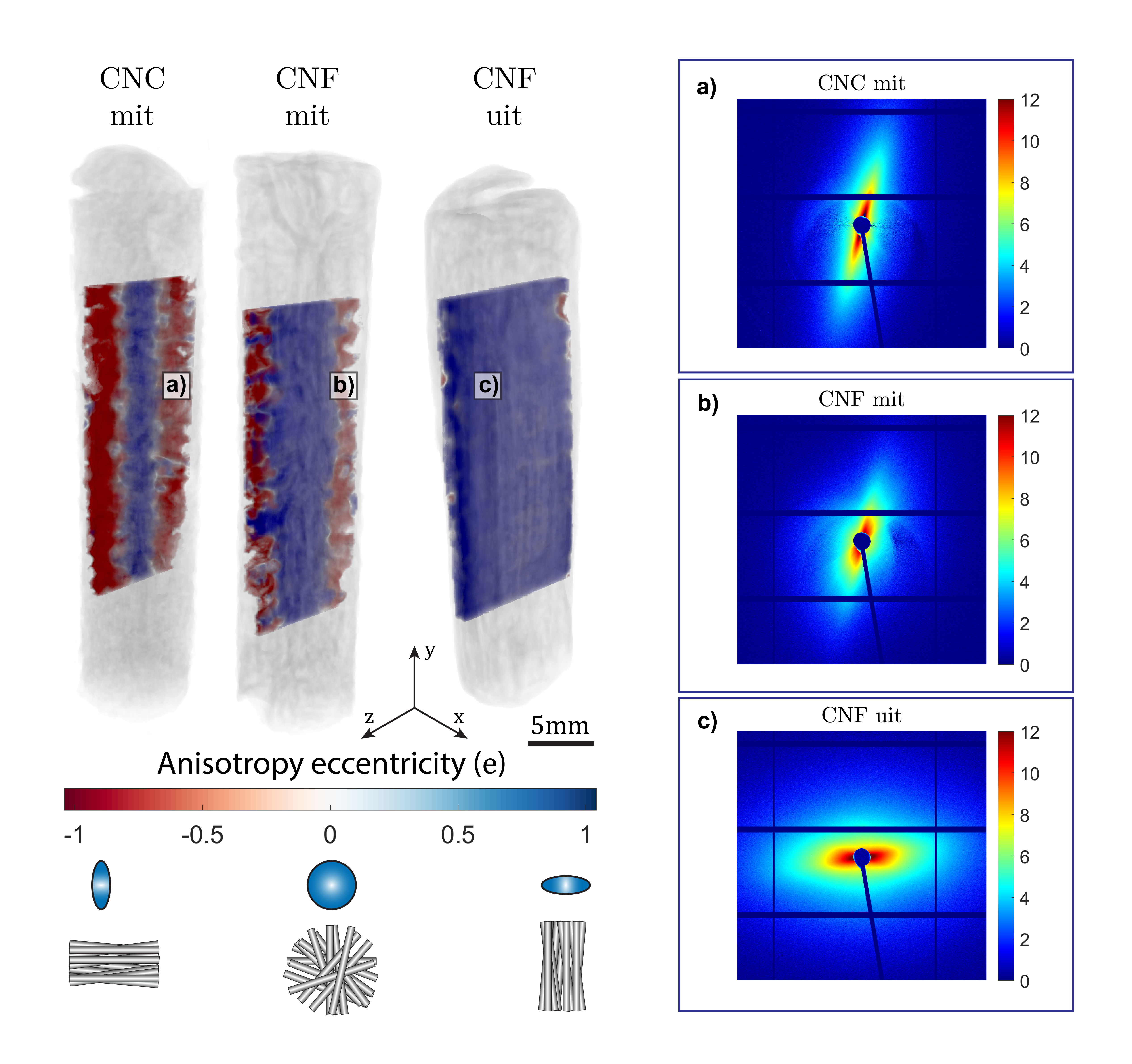}
    \caption{Volumetric reconstructions of the samples undergone multi-directional NDFI tomography. The attenuation contrast of the samples is displayed in gray scale values with transparency adjusted to highlight a thin central section according to the calculated anisotropy eccentricity in each voxel of the samples (in corresponding red to white to blue color map). a) Anisotropic SAXS pattern from the shell region of the CNC mit foam. With reference to the axes reported in the figure, this pattern was collected having the foam aligned such that the incident X-ray beam would be parallel to the y-axis. b) Anisotropic SAXS pattern from the shell region of the CNF mit foam. Also in this case the X-ray beam incident on the sample was aligned along the y-axis. c) Anisotropic SAXS pattern from the central region of the CNF uit foam collected with the X-ray beam aligned along the x-axis.}
    \label{fig:eccSlices}
\end{figure}
In order to compensate for significant density variations when focusing on the eccentricity the directional dark-field values where normalized to their respective maximum.
It has to be noted that while within this study the analyses of anisotropy has been limited to only a few directions most relevant for the investigated material, omni-directional sensitivity can be achieved by employing gratings suited for that purpose\cite{busi2023multi}, which can in principle enable a full scattering tensor tomography\cite{liebi2015nanostructure}.

Fig.~\ref{fig:eccSlices} shows cuts of the reconstruction of conventional attenuation contrast (in gray values) of the three foams, as well as thin eccentricity volume sections overlayed to the images of the respective samples. It is found that in the sample built with the uit procedure, according to the measured anisotropy within the volume, the alignment of the nanostructures in the foam walls displays a clear vertical alignment (i.e., along the y axis with reference to figure \ref{fig:eccSlices}), relatively homogeneously distributed through the full volume of the sample. This is in good agreement with the expected structure of a foam prepared by unidirectional ice-templating. On the other hand, in both the samples built with the mit procedure the alignment of the nanostructure follows a core-shell distribution model, with the center of the sample having the fibrils in the foam walls preferentially aligned vertically, and the shell of the sample having the foam walls preferentially aligned horizontally i.e., radially in the xz-plane (with reference to figure \ref{fig:eccSlices}). This is explained by the alignment dynamics of the multidirectional ice-templating process, where cooling occurs from multiple directions. This method directs the fibers towards the foam’s center, where they are increasingly compacted by the advancing freezing front. As ice crystals grow from opposite directions and converge, the fibers are constrained into tighter spaces and their possibility to align along the initial radial freezing direction diminishes. At these collision points the fibers reorient themselves to fit the diminishing available space, resulting in an axial alignment along the y direction at the foam’s core. Conversely, at the foam’s periphery (the shell), ice crystals tend to solidify more quickly due to their proximity to the cooling source, and far from the central collision axis, encouraging a more radial alignment of the fibers. This effect is accentuated by the lateral growth of ice crystals being more pronounced than vertical growth in these regions. Furthermore, as the ice occupies more space within the foam, the fibers are pressed together, densely packing to maintain volume, ultimately adopting a configuration that optimally accommodates the spatial limitations and directional cues provided by the ice crystal growth. For this reason the core of the mit foams displays a higher density with respect to the shell.

\begin{figure}[h!]
    \centering
    \includegraphics[width=0.9\linewidth]{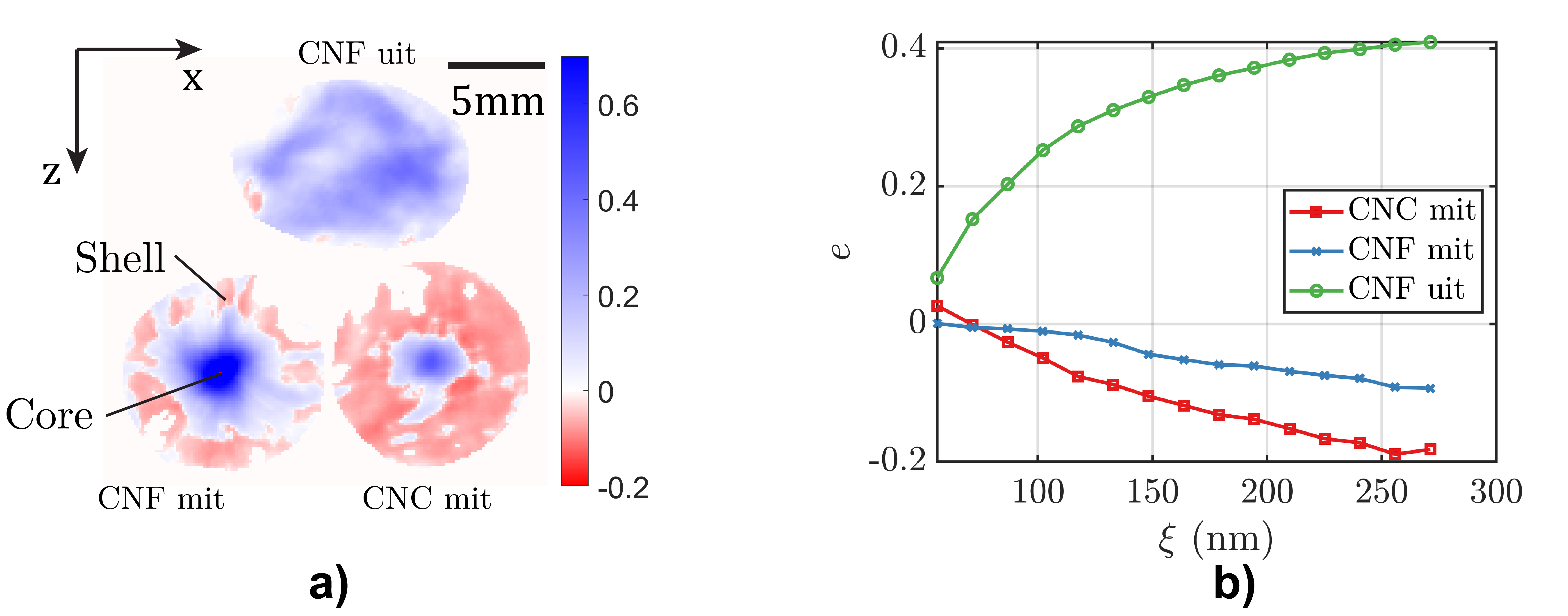}
    \caption{a) Planar slice of the volumetric anisotropy eccentricity of the structures in the three samples. The direction of the cartesian coordinates correspond to the ones displayed in Fig.~\ref{fig:eccSlices}. b) Eccentricity plots as function of autocorrelation length for the three sample types. For the mit samples the region of interest is taken in the shell of the sample whereas for the uit is throughout the full sample bulk.}
    \label{fig:dfplots}
\end{figure}

Fig.\ref{fig:dfplots} a) shows a planar slice of the reconstructed volumetric anisotropy eccentricity where all of the three samples measured are visible for comparisons. The slice view confirms a rather uniform orientation of the nanostructure along the y-axis for the CNF uit sample, and a core/shell structure alignment for the CNF and CNC mit foams. Concerning the samples produced with the mit method, some remarkable differences are observed depending on the nature of the nanocellulose particles used in the foam preparation. The shell part of the CNC sample appears thicker, more uniform, and more aligned in the xz-plane compared to the CNF sample, whereas the core part exhibits stronger vertical alignment with the cylinder axis and higher scattering intensity, thus density, compared to the former. It has to be noted that, since the current formalism does not extend to a full scatter tensor tomography, only radial and vertical alignment are in the focus. Fig.\ref{fig:dfplots} b) displays plots of the anisotropy, in terms of the corresponding eccentricity, as a function of structural length scales, for the bulk of the CNF uit sample and for the isolated shell region of the CNC and CNF mit samples. The CNF uit exhibits a remarkable degree of alignment across the entire probed range, with positive eccentricity values showing a monotonically increasing logarithmic trend as the autocorrelation length $\xi$ increases. This indicates that the alignment of the CNF uit foam becomes more pronounced at larger length scales, thereby implying an increasingly higher degree of structural organization that tends to flatten out above 200 nm. On the other hand, the alignment degree of the shell region of the mit samples also shows an monotonically increasing trend with the values of the eccentricity acquiring more and more negative values as the probed correlation length $\xi$ increases. However, this alignment degree is generally less pronounced with respect to the uit foam, as the absolute values of eccentricity for the mit foams are found to be in closer proximity to zero and the rate of change in the trend of the eccentricity values is lower. Here the alignment degree for the CNC mit sample is overall higher than the one in the CNF mit sample throughout the whole $\xi$ range probed. Similar to the case of the uit foam, the thrend of the eccentricity values towards a radial alignment seems to level out in both the CNC and CNF mit foams from $\xi \sim$ 200 nm and above.

The marked differences between the CNC and CNF mit foams are due to the fact that flexibility and length of the nanocellulose fibers also play a role in how they align and whether they can easily redirect. Being the CNF longer and more flexible with respect to the CNC, CNF are more prone to bend when encountering other fibers, which makes them likely to conform to the available spaces between growing ice crystals during ice-templating. This leads to a more pronounced vertical alignment in the core, as the CNF are pushed in the center of the foam by the ice front, and to a poor radial alignment in the shell, where the constrained radial space hinders the CNF from achieving a uniform and mutually parallel arrangement. On the other hand, due to their stiffness, CNC do not redirect as easily upon encountering the converging ice fronts, leading to less variation in orientation. CNC are also significantly shorter than CNF, therefore the radial spatial constraints do not excessively hinder the shell alignment in the CNC mit foam during ice-templating. This can be clearly seen in Fig.~\ref{fig:eccSlices} and~\ref{fig:dfplots}, where the core region in the CNF mit foam expands to a larger extent with respect to the core region in the CNC mit foam, and their respective shell alignment degrees are very different, despite the freezing rate being equal in the preparation of both samples.

It should be noted that the concept of core/shell alignment within a single sample is not extensively documented, as it represents a complex behavior during ice-templating. Typically, the literature tends to concentrate on unidirectional structures, showcasing uniform alignment across the material, or on gradient structures characterized by a gradual shift in orientation from one side to the other. This focus likely stems from limitations in the currently used investigation methods, which do not allow truly multiscale structural characterizations. The present study on mit nanocellulose foams provides a clear representative case of how the introduction of multi-directional neutron dark-field imaging enhances our ability to study complex hierarchical structures, providing a direct view of phenomena such as core/shell alignment, which cannot be easily characterized in their entirety with other techniques.

\subsection{Cross-Validation with SAXS and SEM Data}
The structural alignment observed in NDFI was validated by a local 2D SAXS measurement for each of the samples built under the same conditions, in parts of the sample denoted by the letters a, b, c in the reconstructed volumes in Fig. ~\ref{fig:eccSlices}, corresponding to the labelling of the respective SAXS patterns in Fig. ~\ref{fig:eccSlices} a-c) (i.e., in the shell region for the mit foams and in a rather central region for the uit foam). SAXS is widely regarded as the standard method for investigating fiber alignment of opaque samples at the nanoscale within volumes delimited by the beam path. Here the samples were positioned in a way that allowed the X-ray beam to interact with them along specific orientations. The uit foam was oriented vertically, so that the beam direction and the foam’s y-axis would be mutually perpendicular, while the mit foams were oriented horizontally, with the beam impinging on the sample perpendicularly to the xz-plane. It should be noted that in order to perform such measurement the samples were cut while still frozen to achieve a size suitable for a SAXS experiment ($\sim$2 mm thickness) while minimizing structural damage. The resulting scattering patterns displayed in Fig.~\ref{fig:eccSlices} a-c exhibit a distinctive anisotropic feature, with a pronounced streak indicating directional scattering, which suggests that there is a preferential orientation of nano-structures in the samples. In order to quantitatively estimate the extent of the fiber orientation in the foams, we calculated their Hermans orientation parameter (HOP) within the full $Q$-range probed in our SAXS experiment (corresponding to a spatial size range from 4 to 150 nm). The HOP is a metric used in small angle scattering analysis to quantify the degree of alignment in fibrous materials. It ranges from -0.5 to 1, where a value of 1 indicates perfect alignment of fibers along a particular reference direction, zero indicates a random, isotropic distribution, and -0.5 indicates perfect alignment perpendicular to the reference direction~\cite{kaniyoor2021quantifying}. The reference direction in our case corresponds to the vertical central axis of the SAXS plots in Fig.~\ref{fig:eccSlices} a-c). The HOP can be considered as the SAXS equivalent of the eccentricity of the correlation function in NDFI (Fig.~\ref{fig:eccSlices}) as both these parameters describe the degree and direction of alignment in the sample (further details on the HOP calculations can be found in the Methods section).

The value of the HOP estimated for the CNC mit foam periphery at the lowest $Q$-point achievable in our measurement conditions (corresponding to $\sim$ 150 nm length scale) is -0.395. This HOP value, close to -0.4, indicates that the bundles of nanofibrils within the shell of the CNC mit sample tend to be relatively well aligned, and predominantly lie along the in-plane direction perpendicular to the reference axis used for the measurement, thereby confirming the alignment direction observed in NDFI. The HOP value being closer to -0.5 than to 0 suggests a relatively strong degree of this preferential perpendicular alignment, albeit not a perfect one. This is not surprising as perfect alignment in the mit process is challenging to achieve due to the complex interplay of the CNC structural properties, processing conditions, and dynamics of the system. Small variations in local conditions within the foam can easily lead to heterogeneities in alignment, which possess nonetheless a high degree of structural consistency and order within the CNC mit foam in our case. 

The value of the HOP estimated for the CNF mit foam is -0.162. This HOP value suggests that the CNF here exhibit a preferential orientation, but this preference is less pronounced than in the CNC foam discussed above. A value closer to 0 indicates a more random orientation compared to the CNC foam, with a slight tendency toward perpendicular alignment relative to the reference axis. This is expected since CNF are longer and more flexible than CNC, which can lead to entanglement and less pronounced alignment during multidirectional ice-templating while CNC, being shorter and more rigid, can align more easily under such radial directional forces. 
The difference in orientation degree between CNF and CNC within mit foams, as reflected in their Hermans Orientation Parameters, is in good agreement with the results of NDFI for the respective regions.

\begin{figure}[h!]
    \centering
    \includegraphics[width=0.9\linewidth]{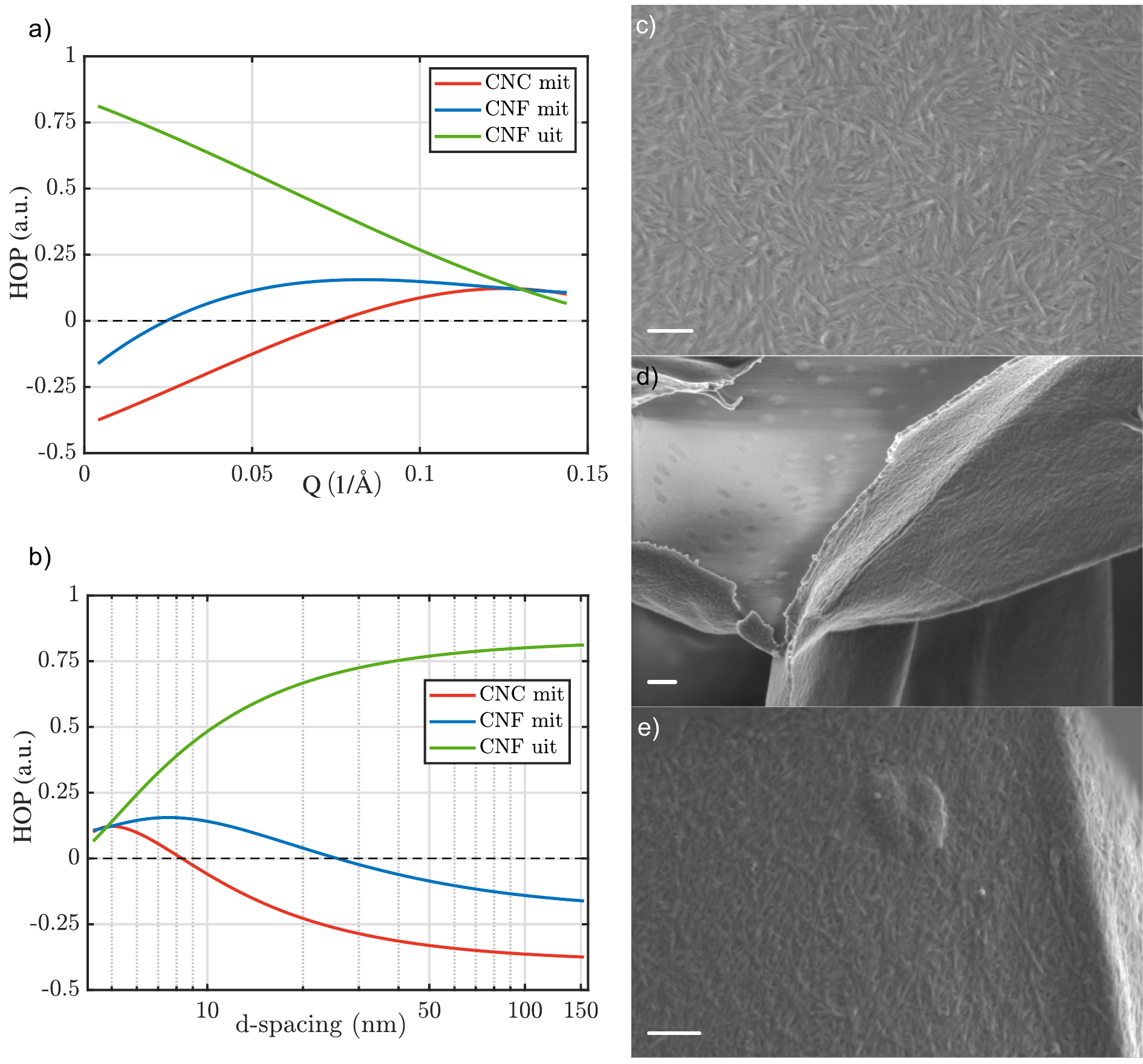}
    \caption{a) Distribution of nanofibril orientation as a function of the scattering vector $Q$ for the CNC mit, the CNF mit and the CNF uit foams. b) Distribution of nanofibril orientation as a function of the length scale in direct space. Scanning electron microscopy (SEM) images of the foam walls in c) CNC mit foam, d) the CNF mit foam, and e) the CNF uit foam. The scale bars for all the SEM images are equivalent to 200 nm.}
    \label{hop}
\end{figure}

The value of the HOP estimated for the rather central region of the CNF uit foam is 0.864, which is indicative of a very high degree of alignment of the CNF within the foam walls, with the orientation being nearly parallel to the sample's cylinder axis in a well-controlled structure. This suggests that the CNF in the uit foam are predominantly aligned in the same direction, which is in stark contrast with the less-aligned CNF in the mit foam, and reflects the different freeze-casting strategies applied for the two samples. The homogeneity of the nanostructures and their alignment direction observed from the SAXS measurements in the CNF uit foams is also in good agreement with the structural features observed in the NDFI measurements. 
The distribution of nanofibril orientation as a function of the scattering vector $Q$ and real space d-spacing is displayed in Fig.~\ref{hop} a,b), along with scanning electron microscopy (SEM) images showing the morphology of the foam walls for the three samples (Fig.~\ref{hop} c-e).

To handle missing data points in the intensity distribution (coming from gaps in the data due to the tiles of the detector), a polynomial interpolation was performed over valid data points, creating a continuous function representing HOP across the entire $Q$ range accessed by our SAXS measurements (Fig.~\ref{hop} a). Fig.~\ref{hop} b) displays the same quantity represented as a function of the length scale in real space for comparison with the length scale dependence of the NDFI eccentricity. All the foams display a strong $Q$-dependence of the orientation parameter: in all cases, as $Q$ increases and the real-space length scale decreases, the orientation becomes less pronounced, with HOP approaching 0, moving towards a more random distribution of fibers at the highest $Q$-point (corresponding to the smallest length scale detectable). Notably, the $Q$-dependence of the HOP for the mit foams overcomes zero and acquires small positive values at around $Q$ = 0.075 Å$^{-1}$ for the CNC (corresponding to $\sim$8.5 nm length scale), and at around $Q$ = 0.024 Å$^{-1}$ for the CNF (corresponding to $\sim$25.6 nm length scale) (Fig.~\ref{hop} a,b).

The change of sign in the $Q$ dependence of the CNC and CNF mit foams is likely due to the fact that, as clearly seen in the NDFI eccentricity maps in Fig. \ref{fig:eccSlices}, the mit samples have regions with different orientation directions with
respect to the reference axis (i.e., parallel for the core and perpendicular for the shell). At large scales, these regions might contribute to an averaged-out orientation that seems more pronounced in SAXS. At smaller scales, the SAXS beam detects more of the variation in local orientation, leading to positive values in the measured orientation parameter. The fraction of positive values for the HOP $Q$-dependence in the CNF mit foam is significantly larger than the HOP values for the CNC mit, as well as the rate of misalignment. In fact, for the CNF mit foam the HOP acquires values closer to zero with respect to the CNC mit HOP, in a broad $Q$ region, indicating a lower degree of preferential orientation for the CNF in the shell region of the mit foam (Fig.~\ref{hop} a,b). These observations support the spatially resolved NDFI results, showing a significantly larger core region for the CNF mit sample with respect to the CNC mit (Fig.~\ref{fig:eccSlices} and~\ref{fig:dfplots} a), as well as an overall lower alignment degree in its shell region (Fig.~\ref{fig:dfplots} b).
The HOP values for the CNF uit foam show a linearly decreasing trend with increasing $Q$, indicating an increasingly random alignment, moving towards a more isotropic structure at smaller length scales. Unlike the mit samples, in the HOP $Q$-dependence for the uit foam the change in sign for the HOP values does not occur throughout the probed $Q$-range (Fig.~\ref{hop} a,b), indicating absence of re-orientation and significant alignment of the fibres with the sample axis across the measured length scale. This is also in good agreement with the results of the NDFI measurements, showing a homogeneously distributed preferential vertical alignment in the full uit foam volume. It should be mentioned that, although both DFI eccentricity and SAXS orientation parameter describe structural orientation, they derive from different analyses due to the different nature of the acquired data with respect to real and reciprocal space, respectively. While the SAXS HOP is extracted from the angular distribution of scattered X-rays, the DFI eccentricity is deduced from directional spatial correlation functions with regards to the dark-field image. Thus, absolute values of the different orientation analyses cannot be compared directly. Despite of these fundamental differences, within the length scale range of relevance for comparison between these two parameters, the cross-correlation of the resulting trends and ratios for the limited regions probed with X-ray scattering is confirming the results of the DFI investigation.

The SEM images (Fig.~\ref{hop} c-e) display a densely packed fibrous structure from the foam walls creating a mat-like texture with bundles of $\sim$40 nm for the CNC foam and $\sim$20 nm for the CNF foams. Unlike slight anisotropic orientation detected by SAXS in this size range, these fiber bundles do not seem to possess a clear orientation in the SEM images. This apparent discrepancy is due to several factors. The orientation of fibers in composites is typically difficult to discern through SEM when fibers are embedded within a matrix or when the contrast between the fibers and the surrounding material does not highlight the alignment clearly. Moreover SEM is a surface technique providing information for a very limited portion of the sample on a relatively large length scale (micrometers). In contrast, SAXS is a bulk method and measures the scattering of X-rays as they interact with structures at the nanoscale, detecting the average orientation and possible anisotropy within the sample volume, even if this is not immediately visible in SEM images. The complex, multiscale nature of these materials, where different structures and behaviors can be observed at different length scales, is clearly exemplified by the results of the SAXS and SEM complementary measurements. The SEM images do not directly show the fiber anisotropy because this structural feature is only detectable at the nanoscale through the bulk of the foam. While not directly showing nano-structural anisotropy, SEM data support the NDFI findings by confirming that the detected orientations are truly reflective of the nanoscale structure.

Similarly to SAXS, NDFI allows to observe overall nano-scale features like phase boundaries, defects, and fiber orientations, not visible in microscopy techniques such as SEM. But, similarly to microscopy and imaging techniques, these features are visible in real space and measurable on large sample volumes and surfaces. This is remarkable since in scattering techniques such as SAXS the actual spatial arrangements of nanostructures can only be inferred indirectly, through data collected in the reciprocal space. The direct observation provided by NDFI simplifies the data interpretation process and provides immediate insights into materials’ internal structure on large scale native samples. In this regard it can be said that NDFI offers the best of both worlds as it bridges the invisible realm of reciprocal space with the tangible reality of the macroscopic visible world. For this reason, the implications of our findings extend far beyond the realm of nanocellulose foams. 

By displaying the wide potential of NDFI through the lens of nanocellulose foam characterization, we lay the groundwork for its application across a wide spectrum of material systems. NDFI's non-invasive nature and ability to provide truly multiscale structural analysis position it as a foundational method for exploring sustainable materials and biological systems with hierarchical structures, thereby facilitating the development of next-generation materials and aiding the industrial upscaling of existing nanotechnologies. On that account, we encourage the scientific community to explore the broader implications of this method for the investigation of diverse systems and mechanisms such as the ones involved in pathological processes (e.g., formation of cancer metastasis~\cite{conceiccao2024unveiling}), in the development of biomimetic materials~\cite{walther2010large}, or in the \textit{in-situ} self-assembly of nanomaterials for tissue engineering~\cite{place2009complexity} and nanoelectronics~\cite{tapio2016toward}, to mention but a few examples. This multi-scale approach to material innovation holds promise for breakthroughs in medical research, bio-inspired materials, and energy-efficient building components.


\section{Methods}
\subsection{Sample preparation}
 
The samples, prepared at the department of Materials and Environmental Chemistry of Stockholm University, consisted of highly oriented nanocellulose foams obtained by ice-templating~\cite{fan2018creating,xu2019efficient} of cellulose nanocrystals (CNC) and cellulose nanofibrils (CNF). While both these cellulose nanomaterials display an elongated structure, the former, short and rigid (average length 150 - 200 nm), are characterized by a higher degree of crystallinity (60$\%$-80$\%$), while the latter, longer and flexible (average length 300 - 500 nm), present a lower degree of crystallinity (45$\%$-60$\%$). 
A total of three foams were prepared: CNC mit, CNF mit and CNF uit.

The CNC mit foam was prepared by dispersing 3 wt$\%$ CNC powder (commercially available via CelluForce Inc.) in deionized water under continuous mechanical stirring. 3 g of such dispersion was transferred to a cylindrical copper mold equipped with a 1 mm thermal pad in the bottom, having diameter 11 mm and 50 mm height. The mold was put in thermal contact with LN$_2$, resulting in two thermal gradients: one in the xz-plane as well as one weaker along the y-axis. Following the direction of the freezing front, the cellulose nanoparticles in suspension in the water self-assemble in the mit radial arrangement.
The CNF mit foam was prepared by dispersing chemically isolated 2,2,6,6-tetramethylpiperidine-1-oxylradical (TEMPO)-oxidized~\cite{saito2004tempo} wood cellulose nanofibrils in deionised water. The CNF dispersion was upconcentrated to 1.5 wt$\%$ using a rotavap and ulthrathoraxed for 10 minutes to ensure a homogenous dispersion. The reuslting dispersion was centrifuged for 5 minutes before ice templating. Here the mit arrangement is obtained with the method described above. To obtain the axial orientation of the fibers, the dispersion was transferred to a Teflon mold equipped with a copper bottom in contact with liquid nitrogen. Following the direction of the ice growth, the cellulose nanofibrils suspended in the dispersion, self-assemble in the uit axial arrangement. 

\subsection{Experimental Methods}
\subsubsection{Neutron Dark Field Imaging}
The NDFI experiments were conducted at the ICON beamline of the Paul Scherrer Institut~\cite{kaestner2011icon}. The samples were measured using a white cold neutron beam incident on a single absorption grating, and a neutron detector comprised of a LiF/ZnS scintillator screen (RC Tritec, Switzerland), a CCD iKon-L camera (2048$\times$2048 pixels) with a magnifying lens, set up such that the active field of view was of 64.5$\times$64.5 mm$^2$ and thus, a pixel size of 31.5 $\upmu$m. The grating used for this measurement was made of a 20 $\upmu$m-thick gadolinium layer patterned into a hexagonal array of 175 $\upmu$m-diameter holes with a period of 350 $\upmu$m. The tomographic reconstructions were carried out for both the transmission and the three directional dark-field imaging modalities enabled by the single absorption grating pattern.
The dark-field in the two diagonal probed directions (60° and 120°) were interpolated to obtain the vertical dark-field whereas the horizontal dark-field was obtained straightforwardly from the directly probed direction.
To reveal the volumetric-resolved anisotropy of the probed microstructure, the eccentricity ($e$) was introduced as:
\[
    e= 
\begin{dcases}
    \sqrt{1-\left(\frac{\mathrm{DF}_{90^\circ}}{\mathrm{DF}_{0^\circ}}\right)^2}, & \text{if  } \mathrm{DF}_{0^\circ} > \mathrm{DF}_{90^\circ} \\
    -\sqrt{1-\left(\frac{\mathrm{DF}_{0^\circ}}{\mathrm{DF}_{90^\circ}}\right)^2}, & \text{if  } \mathrm{DF}_{0^\circ} < \mathrm{DF}_{90^\circ}
\end{dcases}
\]

The samples were wrapped in aluminium foil and placed in an aluminium container on a rotation stage at the sample position. No further manipulation of the foams was conducted.

\subsubsection{Small Angle X-Ray Scattering (SAXS)}
SAXS data were gathered using the P62 instrument at the synchrotron radiation source Petra III (DESY). The X-ray beam dimensions were 200 x 200 $\mu$m, with an energy of 12 keV for the incoming photons. The foams were prepared with a thickness of approximately 2 mm to prevent multiple scattering effects, and the exposure duration for each measurement totaled 1.1 seconds (100 ms times 11 exposures). Notably, given the short exposure time, there were no signs of radiation damage in any of the samples during the experiment. Environmental humidity conditions were controlled in-situ, with a relative humidity level of 50$\%$ RH, matching the environmental humidity conditions of the NDFI experiment. This was achieved using a MHG32 modular humidity generator connected to a die-cast aluminum box sealed with rubber, in which the samples were placed. The box was fitted with two windows covered with Kapton tape for the incoming and scattered X-rays. Dry nitrogen and water vapor were used to achieve the desired humidity level within the box. 
To ascertain the degree of orientation HOP, the 2D anisotropic scattering patterns were azimuthally integrated in the angular range from -179.5$^{\circ}$ to 179.5$^{\circ}$ in steps of 1$^{\circ}$, with a range for the scattering vector $Q$ from 0.00409333 Å$^{-1}$ to 0.143907 Å$^{-1}$ (corresponding to a range in direct space $d = \frac{2\pi}{Q}$ from 153.5 nm to 4.4 nm). Normalization of the scattered intensities to the empty beam and background subtraction were also performed. The resulting normalized intensity distribution was used to calculate Hermans orientation parameter (HOP) for each $Q$-value. HOP is essentially the second order Legendre polynomial $P_2$ evaluated at the mean cosine of the the azimuthal angle $\phi$. This mean value is determined from the angular intensity distribution of scattered X-rays. Therefore, the HOP calculation involved weighting the intensity $I (Q, \phi)$ by the square of the cosine of the azimuthal angle $\phi$, and normalizing this product by the total intensity at each $Q$-value to account for variations in scattering intensity:

\begin{equation}
    \langle P_2 (cos\phi) \rangle = \frac{1}{2} \left(3 \cdot \frac{\sum I_{\phi} cos^2\phi}{\sum I_{\phi}} - 1 \right)
\end{equation}

Legendre polynomials are a series of orthogonal polynomials that are often used as a mathematical tool in physics, when treating problems involving spherical symmetries~\cite{jackson1999classical}. 

\subsubsection{Scanning Electron Microscopy (SEM)}
Scanning electron microscopy images were acquired at the Swiss Federal Laboratories for Materials Science and Technology (EMPA) using a ZEISS GeminiSEM 460 and an inlens secondary electron detector. Due to the insulating nature of the samples, 5 nm carbon film was applied, and the probe current and the acceleration voltage were limited to $\sim$50 pA and $\sim$1kV, respectively. Carbon tape was used to secure the samples on the sample holder. Attempts of measuring the foams with Scanning Transmission Electron Microscopy (STEM) for improved resolution at smaller length scales were unsuccessful due to sensitivity of the samples.

\bibliography{sample}

\begin{thebibliography}{10}
\urlstyle{rm}
\expandafter\ifx\csname url\endcsname\relax
  \def\url#1{\texttt{#1}}\fi
\expandafter\ifx\csname urlprefix\endcsname\relax\def\urlprefix{URL }\fi
\expandafter\ifx\csname doiprefix\endcsname\relax\def\doiprefix{DOI: }\fi
\providecommand{\bibinfo}[2]{#2}
\providecommand{\eprint}[2][]{\url{#2}}

\bibitem{national1994hierarchical}
\bibinfo{author}{Council, N.~R.} \emph{et~al.}
\newblock \emph{\bibinfo{title}{Hierarchical structures in biology as a guide for new materials technology}}, vol. \bibinfo{volume}{464} (\bibinfo{publisher}{National Academies Press}, \bibinfo{year}{1994}).

\bibitem{nechyporchuk2016production}
\bibinfo{author}{Nechyporchuk, O.}, \bibinfo{author}{Belgacem, M.~N.} \& \bibinfo{author}{Bras, J.}
\newblock \bibinfo{journal}{\bibinfo{title}{Production of cellulose nanofibrils: A review of recent advances}}.
\newblock {\emph{\JournalTitle{Industrial Crops and Products}}} \textbf{\bibinfo{volume}{93}}, \bibinfo{pages}{2--25} (\bibinfo{year}{2016}).

\bibitem{yanagi2018intersubband}
\bibinfo{author}{Yanagi, K.} \emph{et~al.}
\newblock \bibinfo{journal}{\bibinfo{title}{Intersubband plasmons in the quantum limit in gated and aligned carbon nanotubes}}.
\newblock {\emph{\JournalTitle{Nature communications}}} \textbf{\bibinfo{volume}{9}}, \bibinfo{pages}{1121} (\bibinfo{year}{2018}).

\bibitem{andersson2017biomimetic}
\bibinfo{author}{Andersson, M.} \emph{et~al.}
\newblock \bibinfo{journal}{\bibinfo{title}{Biomimetic spinning of artificial spider silk from a chimeric minispidroin}}.
\newblock {\emph{\JournalTitle{Nature chemical biology}}} \textbf{\bibinfo{volume}{13}}, \bibinfo{pages}{262--264} (\bibinfo{year}{2017}).

\bibitem{kriechbaum2018analysis}
\bibinfo{author}{Kriechbaum, K.}, \bibinfo{author}{Munier, P.}, \bibinfo{author}{Apostolopoulou-Kalkavoura, V.} \& \bibinfo{author}{Lavoine, N.}
\newblock \bibinfo{journal}{\bibinfo{title}{Analysis of the porous architecture and properties of anisotropic nanocellulose foams: a novel approach to assess the quality of cellulose nanofibrils (cnfs)}}.
\newblock {\emph{\JournalTitle{ACS sustainable chemistry \& engineering}}} \textbf{\bibinfo{volume}{6}}, \bibinfo{pages}{11959--11967} (\bibinfo{year}{2018}).

\bibitem{moon2011cellulose}
\bibinfo{author}{Moon, R.~J.}, \bibinfo{author}{Martini, A.}, \bibinfo{author}{Nairn, J.}, \bibinfo{author}{Simonsen, J.} \& \bibinfo{author}{Youngblood, J.}
\newblock \bibinfo{journal}{\bibinfo{title}{Cellulose nanomaterials review: structure, properties and nanocomposites}}.
\newblock {\emph{\JournalTitle{Chemical Society Reviews}}} \textbf{\bibinfo{volume}{40}}, \bibinfo{pages}{3941--3994} (\bibinfo{year}{2011}).

\bibitem{lavoine2017nanocellulose}
\bibinfo{author}{Lavoine, N.} \& \bibinfo{author}{Bergstr{\"o}m, L.}
\newblock \bibinfo{journal}{\bibinfo{title}{Nanocellulose-based foams and aerogels: Processing, properties, and applications}}.
\newblock {\emph{\JournalTitle{Journal of Materials Chemistry A}}} \textbf{\bibinfo{volume}{5}}, \bibinfo{pages}{16105--16117} (\bibinfo{year}{2017}).

\bibitem{ferreira2020porous}
\bibinfo{author}{Ferreira, F.~V.} \emph{et~al.}
\newblock \bibinfo{journal}{\bibinfo{title}{Porous nanocellulose gels and foams: Breakthrough status in the development of scaffolds for tissue engineering}}.
\newblock {\emph{\JournalTitle{Materials Today}}} \textbf{\bibinfo{volume}{37}}, \bibinfo{pages}{126--141} (\bibinfo{year}{2020}).

\bibitem{chen2023lightweight}
\bibinfo{author}{Chen, C.} \emph{et~al.}
\newblock \bibinfo{journal}{\bibinfo{title}{Lightweight, thermally insulating, fire-proof graphite-cellulose foam}}.
\newblock {\emph{\JournalTitle{Advanced Functional Materials}}} \textbf{\bibinfo{volume}{33}}, \bibinfo{pages}{2204219} (\bibinfo{year}{2023}).

\bibitem{wicklein2015thermally}
\bibinfo{author}{Wicklein, B.} \emph{et~al.}
\newblock \bibinfo{journal}{\bibinfo{title}{Thermally insulating and fire-retardant lightweight anisotropic foams based on nanocellulose and graphene oxide}}.
\newblock {\emph{\JournalTitle{Nature nanotechnology}}} \textbf{\bibinfo{volume}{10}}, \bibinfo{pages}{277--283} (\bibinfo{year}{2015}).

\bibitem{li2023cellulose}
\bibinfo{author}{Li, M.-C.} \emph{et~al.}
\newblock \bibinfo{journal}{\bibinfo{title}{Cellulose nanomaterials in oil and gas industry: Current status and future perspectives}}.
\newblock {\emph{\JournalTitle{Progress in Materials Science}}} \bibinfo{pages}{101187} (\bibinfo{year}{2023}).

\bibitem{apostolopoulou2021thermally}
\bibinfo{author}{Apostolopoulou-Kalkavoura, V.}, \bibinfo{author}{Munier, P.} \& \bibinfo{author}{Bergstr{\"o}m, L.}
\newblock \bibinfo{journal}{\bibinfo{title}{Thermally insulating nanocellulose-based materials}}.
\newblock {\emph{\JournalTitle{Advanced Materials}}} \textbf{\bibinfo{volume}{33}}, \bibinfo{pages}{2001839} (\bibinfo{year}{2021}).

\bibitem{apostolopoulou2021effect}
\bibinfo{author}{Apostolopoulou-Kalkavoura, V.}, \bibinfo{author}{Munier, P.}, \bibinfo{author}{Dlugozima, L.}, \bibinfo{author}{Heuthe, V.-L.} \& \bibinfo{author}{Bergstr{\"o}m, L.}
\newblock \bibinfo{journal}{\bibinfo{title}{Effect of density, phonon scattering and nanoporosity on the thermal conductivity of anisotropic cellulose nanocrystal foams}}.
\newblock {\emph{\JournalTitle{Scientific Reports}}} \textbf{\bibinfo{volume}{11}}, \bibinfo{pages}{18685} (\bibinfo{year}{2021}).

\bibitem{khalil2012green}
\bibinfo{author}{Khalil, H.~A.}, \bibinfo{author}{Bhat, A.} \& \bibinfo{author}{Yusra, A.~I.}
\newblock \bibinfo{journal}{\bibinfo{title}{Green composites from sustainable cellulose nanofibrils: A review}}.
\newblock {\emph{\JournalTitle{Carbohydrate polymers}}} \textbf{\bibinfo{volume}{87}}, \bibinfo{pages}{963--979} (\bibinfo{year}{2012}).

\bibitem{nocerino2022comprehensive}
\bibinfo{author}{Nocerino, E.}
\newblock \emph{\bibinfo{title}{A Comprehensive Experimental Approach to Multifunctional Quantum Materials and their Physical Properties: Geometry and Physics in Condensed Matter}}.
\newblock Ph.D. thesis, \bibinfo{school}{Kungliga Tekniska h{\"o}gskolan} (\bibinfo{year}{2022}).

\bibitem{nocerino2023multiple}
\bibinfo{author}{Nocerino, E.} \emph{et~al.}
\newblock \bibinfo{journal}{\bibinfo{title}{Multiple unconventional charge density wave transitions in lapt2si2 superconductor clarified with high-energy x-ray diffraction}}.
\newblock {\emph{\JournalTitle{Communications Materials}}} \textbf{\bibinfo{volume}{4}}, \bibinfo{pages}{77} (\bibinfo{year}{2023}).

\bibitem{nocerino2023competition}
\bibinfo{author}{Nocerino, E.} \emph{et~al.}
\newblock \bibinfo{journal}{\bibinfo{title}{Competition between magnetic interactions and structural instabilities leading to itinerant frustration in the triangular lattice antiferromagnet licrse2}}.
\newblock {\emph{\JournalTitle{Communications Materials}}} \textbf{\bibinfo{volume}{4}}, \bibinfo{pages}{81} (\bibinfo{year}{2023}).

\bibitem{nocerino2023q}
\bibinfo{author}{Nocerino, E.} \emph{et~al.}
\newblock \bibinfo{journal}{\bibinfo{title}{Q-dependent electron-phonon coupling induced phonon softening and non-conventional critical behavior in the cdw superconductor lapt2si2}}.
\newblock {\emph{\JournalTitle{Journal of Science: Advanced Materials and Devices}}} \textbf{\bibinfo{volume}{8}}, \bibinfo{pages}{100621} (\bibinfo{year}{2023}).

\bibitem{nocerino2024ion}
\bibinfo{author}{Nocerino, E.} \emph{et~al.}
\newblock \bibinfo{journal}{\bibinfo{title}{Na-ion dynamics in the solid solution na x ca 1- x cr 2 o 4 studied by muon spin rotation and neutron diffraction}}.
\newblock {\emph{\JournalTitle{Sustainable Energy \& Fuels}}} \textbf{\bibinfo{volume}{8}}, \bibinfo{pages}{1424--1437} (\bibinfo{year}{2024}).

\bibitem{goh2014directional}
\bibinfo{author}{Goh, P.}, \bibinfo{author}{Ismail, A.} \& \bibinfo{author}{Ng, B.}
\newblock \bibinfo{journal}{\bibinfo{title}{Directional alignment of carbon nanotubes in polymer matrices: Contemporary approaches and future advances}}.
\newblock {\emph{\JournalTitle{Composites Part A: Applied Science and Manufacturing}}} \textbf{\bibinfo{volume}{56}}, \bibinfo{pages}{103--126} (\bibinfo{year}{2014}).

\bibitem{zhang2005aligned}
\bibinfo{author}{Zhang, H.} \emph{et~al.}
\newblock \bibinfo{journal}{\bibinfo{title}{Aligned two-and three-dimensional structures by directional freezing of polymers and nanoparticles}}.
\newblock {\emph{\JournalTitle{Nature materials}}} \textbf{\bibinfo{volume}{4}}, \bibinfo{pages}{787--793} (\bibinfo{year}{2005}).

\bibitem{griffin2022scalable}
\bibinfo{author}{Griffin, A.} \emph{et~al.}
\newblock \bibinfo{journal}{\bibinfo{title}{Scalable methods for directional assembly of fillers in polymer composites: Creating pathways for improving material properties}}.
\newblock {\emph{\JournalTitle{Polymer Composites}}} \textbf{\bibinfo{volume}{43}}, \bibinfo{pages}{5747--5766} (\bibinfo{year}{2022}).

\bibitem{nygaard2024formax}
\bibinfo{author}{Nyg{\aa}rd, K.} \emph{et~al.}
\newblock \bibinfo{journal}{\bibinfo{title}{Formax--a beamline for multiscale and multimodal structural characterization of hierarchical materials}}.
\newblock {\emph{\JournalTitle{Journal of Synchrotron Radiation}}} \textbf{\bibinfo{volume}{31}} (\bibinfo{year}{2024}).

\bibitem{liebi2015nanostructure}
\bibinfo{author}{Liebi, M.} \emph{et~al.}
\newblock \bibinfo{journal}{\bibinfo{title}{Nanostructure surveys of macroscopic specimens by small-angle scattering tensor tomography}}.
\newblock {\emph{\JournalTitle{Nature}}} \textbf{\bibinfo{volume}{527}}, \bibinfo{pages}{349--352} (\bibinfo{year}{2015}).

\bibitem{sanchez2021x}
\bibinfo{author}{Sanchez-Cano, C.} \emph{et~al.}
\newblock \bibinfo{journal}{\bibinfo{title}{X-ray-based techniques to study the nano--bio interface}}.
\newblock {\emph{\JournalTitle{ACS nano}}} \textbf{\bibinfo{volume}{15}}, \bibinfo{pages}{3754--3807} (\bibinfo{year}{2021}).

\bibitem{azevedo2024connectomic}
\bibinfo{author}{Azevedo, A.} \emph{et~al.}
\newblock \bibinfo{journal}{\bibinfo{title}{Connectomic reconstruction of a female drosophila ventral nerve cord}}.
\newblock {\emph{\JournalTitle{Nature}}} \bibinfo{pages}{1--9} (\bibinfo{year}{2024}).

\bibitem{peddie2022volume}
\bibinfo{author}{Peddie, C.~J.} \emph{et~al.}
\newblock \bibinfo{journal}{\bibinfo{title}{Volume electron microscopy}}.
\newblock {\emph{\JournalTitle{Nature Reviews Methods Primers}}} \textbf{\bibinfo{volume}{2}}, \bibinfo{pages}{51} (\bibinfo{year}{2022}).

\bibitem{kagias2019diffractive}
\bibinfo{author}{Kagias, M.} \emph{et~al.}
\newblock \bibinfo{journal}{\bibinfo{title}{Diffractive small angle x-ray scattering imaging for anisotropic structures}}.
\newblock {\emph{\JournalTitle{Nature communications}}} \textbf{\bibinfo{volume}{10}}, \bibinfo{pages}{5130} (\bibinfo{year}{2019}).

\bibitem{wirtensohn2024nanoscale}
\bibinfo{author}{Wirtensohn, S.} \emph{et~al.}
\newblock \bibinfo{journal}{\bibinfo{title}{Nanoscale dark-field imaging in full-field transmission x-ray microscopy}}.
\newblock {\emph{\JournalTitle{Optica}}} \textbf{\bibinfo{volume}{11}}, \bibinfo{pages}{852--859} (\bibinfo{year}{2024}).

\bibitem{strobl2008neutron}
\bibinfo{author}{Strobl, M.} \emph{et~al.}
\newblock \bibinfo{journal}{\bibinfo{title}{Neutron dark-field tomography}}.
\newblock {\emph{\JournalTitle{Physical review letters}}} \textbf{\bibinfo{volume}{101}}, \bibinfo{pages}{123902} (\bibinfo{year}{2008}).

\bibitem{strobl2014general}
\bibinfo{author}{Strobl, M.}
\newblock \bibinfo{journal}{\bibinfo{title}{General solution for quantitative dark-field contrast imaging with grating interferometers}}.
\newblock {\emph{\JournalTitle{Scientific reports}}} \textbf{\bibinfo{volume}{4}}, \bibinfo{pages}{1--6} (\bibinfo{year}{2014}).

\bibitem{strobl2015quantitative}
\bibinfo{author}{Strobl, M.} \emph{et~al.}
\newblock \bibinfo{journal}{\bibinfo{title}{Quantitative neutron dark-field imaging through spin-echo interferometry}}.
\newblock {\emph{\JournalTitle{Scientific reports}}} \textbf{\bibinfo{volume}{5}}, \bibinfo{pages}{1--6} (\bibinfo{year}{2015}).

\bibitem{strobl2017small}
\bibinfo{author}{Strobl, M.}, \bibinfo{author}{Harti, R.~P.}, \bibinfo{author}{Gr{\"u}nzweig, C.}, \bibinfo{author}{Woracek, R.} \& \bibinfo{author}{Plomp, J.}
\newblock \bibinfo{journal}{\bibinfo{title}{Small angle scattering in neutron imaging—a review}}.
\newblock {\emph{\JournalTitle{Journal of Imaging}}} \textbf{\bibinfo{volume}{3}}, \bibinfo{pages}{64} (\bibinfo{year}{2017}).

\bibitem{busi2023multi}
\bibinfo{author}{Busi, M.} \emph{et~al.}
\newblock \bibinfo{journal}{\bibinfo{title}{Multi-directional neutron dark-field imaging with single absorption grating}}.
\newblock {\emph{\JournalTitle{Scientific Reports}}} \textbf{\bibinfo{volume}{13}}, \bibinfo{pages}{15274} (\bibinfo{year}{2023}).

\bibitem{valsecchi2019visualization}
\bibinfo{author}{Valsecchi, J.} \emph{et~al.}
\newblock \bibinfo{journal}{\bibinfo{title}{Visualization and quantification of inhomogeneous and anisotropic magnetic fields by polarized neutron grating interferometry}}.
\newblock {\emph{\JournalTitle{Nature communications}}} \textbf{\bibinfo{volume}{10}}, \bibinfo{pages}{1--9} (\bibinfo{year}{2019}).

\bibitem{valsecchi2020characterization}
\bibinfo{author}{Valsecchi, J.} \emph{et~al.}
\newblock \bibinfo{journal}{\bibinfo{title}{{Characterization of oriented microstructures through anisotropic small-angle scattering by 2D neutron dark-field imaging}}}.
\newblock {\emph{\JournalTitle{Communications Physics}}} \textbf{\bibinfo{volume}{3}}, \bibinfo{pages}{1--8} (\bibinfo{year}{2020}).

\bibitem{howells2009assessment}
\bibinfo{author}{Howells, M.~R.} \emph{et~al.}
\newblock \bibinfo{journal}{\bibinfo{title}{An assessment of the resolution limitation due to radiation-damage in x-ray diffraction microscopy}}.
\newblock {\emph{\JournalTitle{Journal of electron spectroscopy and related phenomena}}} \textbf{\bibinfo{volume}{170}}, \bibinfo{pages}{4--12} (\bibinfo{year}{2009}).

\bibitem{beetz2003soft}
\bibinfo{author}{Beetz, T.} \& \bibinfo{author}{Jacobsen, C.}
\newblock \bibinfo{journal}{\bibinfo{title}{Soft x-ray radiation-damage studies in pmma using a cryo-stxm}}.
\newblock {\emph{\JournalTitle{Journal of Synchrotron Radiation}}} \textbf{\bibinfo{volume}{10}}, \bibinfo{pages}{280--283} (\bibinfo{year}{2003}).

\bibitem{holton2009beginner}
\bibinfo{author}{Holton, J.~M.}
\newblock \bibinfo{journal}{\bibinfo{title}{A beginner's guide to radiation damage}}.
\newblock {\emph{\JournalTitle{Journal of synchrotron radiation}}} \textbf{\bibinfo{volume}{16}}, \bibinfo{pages}{133--142} (\bibinfo{year}{2009}).

\bibitem{olsson2022multiscale}
\bibinfo{author}{Olsson, M.}
\newblock \emph{\bibinfo{title}{Multiscale X-ray Characterisation of Cellulose-based Solid Dispersions}} (\bibinfo{publisher}{Chalmers Tekniska Hogskola (Sweden)}, \bibinfo{year}{2022}).

\bibitem{meents2010origin}
\bibinfo{author}{Meents, A.}, \bibinfo{author}{Gutmann, S.}, \bibinfo{author}{Wagner, A.} \& \bibinfo{author}{Schulze-Briese, C.}
\newblock \bibinfo{journal}{\bibinfo{title}{Origin and temperature dependence of radiation damage in biological samples at cryogenic temperatures}}.
\newblock {\emph{\JournalTitle{Proceedings of the National Academy of Sciences}}} \textbf{\bibinfo{volume}{107}}, \bibinfo{pages}{1094--1099} (\bibinfo{year}{2010}).

\bibitem{glaeser1978radiation}
\bibinfo{author}{Glaeser, R.~M.} \& \bibinfo{author}{Taylor, K.~A.}
\newblock \bibinfo{journal}{\bibinfo{title}{Radiation damage relative to transmission electron microscopy of biological specimens at low temperature: a review}}.
\newblock {\emph{\JournalTitle{Journal of microscopy}}} \textbf{\bibinfo{volume}{112}}, \bibinfo{pages}{127--138} (\bibinfo{year}{1978}).

\bibitem{glaeser1971limitations}
\bibinfo{author}{Glaeser, R.~M.}
\newblock \bibinfo{journal}{\bibinfo{title}{Limitations to significant information in biological electron microscopy as a result of radiation damage}}.
\newblock {\emph{\JournalTitle{Journal of ultrastructure research}}} \textbf{\bibinfo{volume}{36}}, \bibinfo{pages}{466--482} (\bibinfo{year}{1971}).

\bibitem{simons2015dark}
\bibinfo{author}{Simons, H.} \emph{et~al.}
\newblock \bibinfo{journal}{\bibinfo{title}{Dark-field x-ray microscopy for multiscale structural characterization}}.
\newblock {\emph{\JournalTitle{Nature communications}}} \textbf{\bibinfo{volume}{6}}, \bibinfo{pages}{6098} (\bibinfo{year}{2015}).

\bibitem{guo2024grating}
\bibinfo{author}{Guo, P.} \emph{et~al.}
\newblock \bibinfo{journal}{\bibinfo{title}{Grating-based x-ray dark-field ct for lung cancer diagnosis in mice}}.
\newblock {\emph{\JournalTitle{European Radiology Experimental}}} \textbf{\bibinfo{volume}{8}}, \bibinfo{pages}{12} (\bibinfo{year}{2024}).

\bibitem{duncan2022x}
\bibinfo{author}{Duncan, K.~E.}, \bibinfo{author}{Czymmek, K.~J.}, \bibinfo{author}{Jiang, N.}, \bibinfo{author}{Thies, A.~C.} \& \bibinfo{author}{Topp, C.~N.}
\newblock \bibinfo{journal}{\bibinfo{title}{X-ray microscopy enables multiscale high-resolution 3d imaging of plant cells, tissues, and organs}}.
\newblock {\emph{\JournalTitle{Plant Physiology}}} \textbf{\bibinfo{volume}{188}}, \bibinfo{pages}{831--845} (\bibinfo{year}{2022}).

\bibitem{shao2020freeze}
\bibinfo{author}{Shao, G.}, \bibinfo{author}{Hanaor, D.~A.}, \bibinfo{author}{Shen, X.} \& \bibinfo{author}{Gurlo, A.}
\newblock \bibinfo{journal}{\bibinfo{title}{Freeze casting: from low-dimensional building blocks to aligned porous structures—a review of novel materials, methods, and applications}}.
\newblock {\emph{\JournalTitle{Advanced Materials}}} \textbf{\bibinfo{volume}{32}}, \bibinfo{pages}{1907176} (\bibinfo{year}{2020}).

\bibitem{fan2018creating}
\bibinfo{author}{Fan, L.}, \bibinfo{author}{Li, J.-L.}, \bibinfo{author}{Cai, Z.} \& \bibinfo{author}{Wang, X.}
\newblock \bibinfo{journal}{\bibinfo{title}{Creating biomimetic anisotropic architectures with co-aligned nanofibers and macrochannels by manipulating ice crystallization}}.
\newblock {\emph{\JournalTitle{ACS nano}}} \textbf{\bibinfo{volume}{12}}, \bibinfo{pages}{5780--5790} (\bibinfo{year}{2018}).

\bibitem{xu2019efficient}
\bibinfo{author}{Xu, W.} \emph{et~al.}
\newblock \bibinfo{journal}{\bibinfo{title}{Efficient water transport and solar steam generation via radially, hierarchically structured aerogels}}.
\newblock {\emph{\JournalTitle{Acs Nano}}} \textbf{\bibinfo{volume}{13}}, \bibinfo{pages}{7930--7938} (\bibinfo{year}{2019}).

\bibitem{pfeiffer2006neutron}
\bibinfo{author}{Pfeiffer, F.} \emph{et~al.}
\newblock \bibinfo{journal}{\bibinfo{title}{Neutron phase imaging and tomography}}.
\newblock {\emph{\JournalTitle{Physical review letters}}} \textbf{\bibinfo{volume}{96}}, \bibinfo{pages}{215505} (\bibinfo{year}{2006}).

\bibitem{strobl2019achromatic}
\bibinfo{author}{Strobl, M.} \emph{et~al.}
\newblock \bibinfo{journal}{\bibinfo{title}{Achromatic non-interferometric single grating neutron dark-field imaging}}.
\newblock {\emph{\JournalTitle{Scientific Reports}}} \textbf{\bibinfo{volume}{9}}, \bibinfo{pages}{1--7} (\bibinfo{year}{2019}).

\bibitem{busi2022cold}
\bibinfo{author}{Busi, M.}, \bibinfo{author}{Zdora, M.-C.}, \bibinfo{author}{Valsecchi, J.}, \bibinfo{author}{Bacak, M.} \& \bibinfo{author}{Strobl, M.}
\newblock \bibinfo{journal}{\bibinfo{title}{Cold and thermal neutron single grating dark-field imaging extended to an inverse pattern regime}}.
\newblock {\emph{\JournalTitle{Applied Sciences}}} \textbf{\bibinfo{volume}{12}}, \bibinfo{pages}{2798} (\bibinfo{year}{2022}).

\bibitem{vontobel2006neutron}
\bibinfo{author}{Vontobel, P.}, \bibinfo{author}{Lehmann, E.~H.}, \bibinfo{author}{Hassanein, R.} \& \bibinfo{author}{Frei, G.}
\newblock \bibinfo{journal}{\bibinfo{title}{Neutron tomography: Method and applications}}.
\newblock {\emph{\JournalTitle{Physica B: Condensed Matter}}} \textbf{\bibinfo{volume}{385}}, \bibinfo{pages}{475--480} (\bibinfo{year}{2006}).

\bibitem{palenstijn2015distributed}
\bibinfo{author}{Palenstijn, W.~J.}, \bibinfo{author}{B{\'e}dorf, J.} \& \bibinfo{author}{Batenburg, K.~J.}
\newblock \bibinfo{journal}{\bibinfo{title}{A distributed sirt implementation for the astra toolbox}}.
\newblock {\emph{\JournalTitle{Proc. Fully Three-Dimensional Image Reconstruct. Radiol. Nucl. Med}}} \bibinfo{pages}{166--169} (\bibinfo{year}{2015}).

\bibitem{van2016fast}
\bibinfo{author}{Van~Aarle, W.} \emph{et~al.}
\newblock \bibinfo{journal}{\bibinfo{title}{Fast and flexible x-ray tomography using the astra toolbox}}.
\newblock {\emph{\JournalTitle{Optics express}}} \textbf{\bibinfo{volume}{24}}, \bibinfo{pages}{25129--25147} (\bibinfo{year}{2016}).

\bibitem{kaniyoor2021quantifying}
\bibinfo{author}{Kaniyoor, A.}, \bibinfo{author}{Gspann, T.~S.}, \bibinfo{author}{Mizen, J.~E.} \& \bibinfo{author}{Elliott, J.~A.}
\newblock \bibinfo{journal}{\bibinfo{title}{Quantifying alignment in carbon nanotube yarns and similar two-dimensional anisotropic systems}}.
\newblock {\emph{\JournalTitle{Journal of Applied Polymer Science}}} \textbf{\bibinfo{volume}{138}}, \bibinfo{pages}{50939} (\bibinfo{year}{2021}).

\bibitem{conceiccao2024unveiling}
\bibinfo{author}{Concei{\c{c}}{\~a}o, A.~L.} \emph{et~al.}
\newblock \bibinfo{journal}{\bibinfo{title}{Unveiling breast cancer metastasis through an advanced x-ray imaging approach}}.
\newblock {\emph{\JournalTitle{Scientific Reports}}} \textbf{\bibinfo{volume}{14}}, \bibinfo{pages}{1448} (\bibinfo{year}{2024}).

\bibitem{walther2010large}
\bibinfo{author}{Walther, A.} \emph{et~al.}
\newblock \bibinfo{journal}{\bibinfo{title}{Large-area, lightweight and thick biomimetic composites with superior material properties via fast, economic, and green pathways}}.
\newblock {\emph{\JournalTitle{Nano letters}}} \textbf{\bibinfo{volume}{10}}, \bibinfo{pages}{2742--2748} (\bibinfo{year}{2010}).

\bibitem{place2009complexity}
\bibinfo{author}{Place, E.~S.}, \bibinfo{author}{Evans, N.~D.} \& \bibinfo{author}{Stevens, M.~M.}
\newblock \bibinfo{journal}{\bibinfo{title}{Complexity in biomaterials for tissue engineering}}.
\newblock {\emph{\JournalTitle{Nature materials}}} \textbf{\bibinfo{volume}{8}}, \bibinfo{pages}{457--470} (\bibinfo{year}{2009}).

\bibitem{tapio2016toward}
\bibinfo{author}{Tapio, K.} \emph{et~al.}
\newblock \bibinfo{journal}{\bibinfo{title}{Toward single electron nanoelectronics using self-assembled dna structure}}.
\newblock {\emph{\JournalTitle{Nano letters}}} \textbf{\bibinfo{volume}{16}}, \bibinfo{pages}{6780--6786} (\bibinfo{year}{2016}).

\bibitem{saito2004tempo}
\bibinfo{author}{Saito, T.} \& \bibinfo{author}{Isogai, A.}
\newblock \bibinfo{journal}{\bibinfo{title}{Tempo-mediated oxidation of native cellulose. the effect of oxidation conditions on chemical and crystal structures of the water-insoluble fractions}}.
\newblock {\emph{\JournalTitle{Biomacromolecules}}} \textbf{\bibinfo{volume}{5}}, \bibinfo{pages}{1983--1989} (\bibinfo{year}{2004}).

\bibitem{kaestner2011icon}
\bibinfo{author}{Kaestner, A.} \emph{et~al.}
\newblock \bibinfo{journal}{\bibinfo{title}{{The ICON beamline--A facility for cold neutron imaging at SINQ}}}.
\newblock {\emph{\JournalTitle{Nuclear Instruments and Methods in Physics Research Section A: Accelerators, Spectrometers, Detectors and Associated Equipment}}} \textbf{\bibinfo{volume}{659}}, \bibinfo{pages}{387--393} (\bibinfo{year}{2011}).

\bibitem{jackson1999classical}
\bibinfo{author}{Jackson, J.~D.} \& \bibinfo{author}{Fox, R.~F.}
\newblock \bibinfo{title}{Classical electrodynamics} (\bibinfo{year}{1999}).

\end{thebibliography}

\section*{Acknowledgements}
E.N. acknowledges financial support from the SSF-Swedness grant SNP21-0004 and the Foundation Blanceflor 2024 fellow scholarship. A.Å. acknowledges the SSF funded graduate school SwedNess (grant number: GSn15-008). We acknowledge DESY (Hamburg, Germany), a member of the Helmholtz Association HGF, for the provision of experimental facilities. Parts of this research were carried out at PETRA III, beamtime was allocated for proposal I-20230270 EC, and we would like to sincerely thank Dr. Andre Luiz Coelho Conceicao for assistance in using P62. The authors wish to thank Dr. Sina Azad of the Swiss Federal Laboratories for Materials Science and Technology (EMPA) for producing the SEM images attached to this work.

\section*{Author contributions statement}
M.B., conceptualization, NDFI data collection and analysis; E.N., conceptualization and funding acquisition, SAXS data collection and analysis; AÅ, SAXS data collection and sample preparation; L.B., conceptualization and resources; M.S., conceptualization and resources. E.N. and M.B. produced the original and final draft and all co-authors reviewed/edited the final draft.

\section*{Data availability}
All the data are available from the corresponding authors upon reasonable request.

\end{document}